\documentclass[12pt,a4paper]{article}

\usepackage{graphics}

\begin{document}

\title{QUANTUM PROCESSES AND THE FOUNDATION OF RELATIONAL THEORIES OF SPACE AND TIME}
\author{Miguel Lorente \\
Departamento de F{\'\i}sica. Facultad de Ciencias. \\
Universidad de Oviedo. E-33007 Oviedo, Spain.}
\date{}

\maketitle

\begin{abstract}
We present current theories about the structure of space and time, where the
building blocks are some fundamental entities (yes-no experiment, quantum
processes, spin net-work, preparticles) that do not presuppose the existence of
space and time. The relations among these objects are the base for a pregeometry
of discrete character, the continuous limit of which gives rise to the physical
properties of the space and time.
\end{abstract}

\section{Introduction}

The theory about the structure of space and time has been discussed from very
long time not only among philosophers but also among physicists. It is very well
known the correspondence between Newton's disciple S. Clark and Leibniz about the
absolute or relational character of space and time \cite{1}. The laws of
mechanics and electromagnetism were written with the hypothesis of an absolute
space and time but relational theories were unable to construct a consistent
model of physical world and were abandoned. With the invention of the special
theory of relativity the old arguments of relational theories were renewed but
only in the phenomenological level. Einstein himself recognized the atractive
ideas of Leibniz as a pioneering work in the theory of relativity \cite{2}.

In this century the controversy among the physicists who defend the absolute or
relational theory of space and time has increased in great amount from different
reasons \cite{3}. The experimental and theoretical development of general
relativity and cosmological models have forced to review the concepts of space
and time. The revision of the foundations of quantum mechanics has moved some
physicists to reduce the structure of space and time to some properties of
quantum processes \cite{4}. Finally, the study of fundamental axioms of geometry
(in the mathematical as well in the physical sense) has introduced new light
into the discussion about the primitive or derived character of space and time.

\section{Some modern theories about the relational character of space and time}

We don't want to go again in the arguments given by the two positions. Instead
we will expand some recent relational theories of space and time to prove that
the basic assumption common to all of them is that there exists some fundamental
objects out of them the structure of space and time is constructed. We will
identify those objects as events, processes, yes-no experiments, monads
(following Leibniz's ideas) and so on. Therefore the conceps of space and time
are for these authors derived concepts in the logical as well in the ontological
level.

\medskip
{\sc R. Penrose} \cite{5}. His model starts with objects and the interrelations
between objects. An object is thus ``located'' either directionally or
positionally in terms of its relations with other objects. ``One does not really
need a space to begin with. The notion of space comes out as a convenience at
the end''. The physical objects consist of some units with definite spin acting
among themselves giving rise to a N-unit with a total spin representing a
direction. The interaction of two N-units es interpreted as the angle between
the two directions.

\medskip
{\sc V. Kaplunovsky, M. Weinstein} \cite{6}. They consider the notion of
space-time continuum as an illusion of low energy dynamics. Quantum systems are
defined without the notion of space. The $x_m$ variables have to be replaced by
the labels for quantum degrees of freedom. Bose and Fermi fields satisfy the
canonical commutation relation, where the quantum degree of freedom play the
role of site variable.

$$\left[{\Pi \left(\ell \right), \phi \left(m \right)}\right]
=-i\delta_{\ell m},\left\{\Psi_{\alpha }^+ \left(\ell\right),\Psi_\beta
\left(m\right)\right\}=\delta_{\alpha \rho}\delta_{\ell m}$$

The hamiltonian incorporates link fields $X_{(\ell m)}$ and $P_{(\ell m)}$ to boson and
fermion fields, satisfying

$$\left[P_{(\ell m)}^{\alpha \beta} X_{(\ell 'm')}^{\alpha '\beta
 '}\right]=-i\delta _{\ell \ell '}\delta_{mm'}\delta
_{\alpha \alpha '}\delta_{\beta \beta '}$$

The hamiltonian is constructed out of such fields.

$$H=\sum\nolimits\limits_{p,q}^{} \left[-iX_{(p'p)}\Psi ^+ {(p',q)} \sigma_x
\Psi {(p,q)} - i{X}_{(q'q)}\Psi ^+ {(p,q')}\sigma_y\Psi {(p,q)}\right]$$

All the fields are labelled by integer indexes, that is, they are connected
among themselves as a finite network, with the structure of 1-simplex, (where
each field is acting with all the other fields) and the different 1-simplices
are interacting with other 1-simplices giving rise to N-simplex of more
complicated structure. This is called a generalized SLAC lattice, which
underlies the continuum space-time.

\medskip
{\sc M. Garcia Sucre} \cite{7}. This author, after classifying the relational
theories of space and time, proposes a set-theoretical model of space and time
in which the primitive concepts are preparticles as the most basic components of
a physical system. Using the membership relation of set theory he construct the
elementary particles as subsets of the power set of preparticles. Inclusion
relations among preparticles and interaction among them gives rise to the
temporal and spatial structure of the world out of which reference frames and
motions are described.

\medskip
{\sc M. Bunge} \cite{8} proposes a relational theory of physical space. The
world is constituded by things such as elementary particles and fields and
physical systems are composed out of these things. The concept of space is a
derived one, therefore these things are not located in the space. These things
are acting among themselves, the result of these interactions are called
processes or events. Space is nothing more that the collections of these
individual things and their relations.

\medskip
{\sc M. Friedman} \cite{9} starts from the set of concrete physical events ${\cal E}$,
or, the set of space-time points that are actually occupied by material objects
or processes. There is a relationalist's ontology by which the (3+1)-dimensional
continuum is nothing more than a construction out of the set ${\cal E}$ of physical
events.

\medskip
{\sc Mundy} \cite{10} offers a relational theory of Minkowski space-time. Suppose
that there are a finite number of point particles and the mutual spatial relations
among the particles obey the axioms of euclidean geometry. Any talk about space is
to be analyzed according to the representations of the interparticle relations. Of
course these theories must recover the field-theoretical models out of the
individual events that are finite in number. A real challenge for the relational
theories.

\medskip
{\sc J.A. Wheeler} \cite{11} has introduced the concept of pregeometry to
understand the laws of euclidean and non-euclidean geometry. The objects in the
pregeometry are logical in character. Space and time are not physical entities
but conceptions created by man, to keep track of the order of things. Time and
space must be derived in the classical limit from the universe of
our discrete elements (or bits) of information.

\medskip
{\sc Ponzano and Regge} \cite{12} have develope Wheeler's ideas in some model of
curved space time using spin-network, the elementary structure of which are the
$3j$-simbols for 3 particles of spin 1/2 acting among themselves. The structure of
a curve 2-dimensional surface is described by the closed interaction of
half-integral spin particles whose diagramatic representation is given by a
closed finely triangulated 2-surface with $2n$ triangles and $3n$ edges. The
fineness of the diagram, determined by $n$, can be used to derive a
correspondence law between the basic pregeometry and the continuous limit of physical 
geometry. The Ponzano-Regge model has been developped by modern authors in the study 
of Quantum Gravity.

\section{An axiomatic formulation of a euclidean lattice}

In the last section we have reviewed some modern theories in which the concepts of
space and time are derived from the relations among fundamental objects
(building blocks). The underlying structure should be compared with the
properties of the physical space and time. We want to develope now some
pregeometry in which the axioms are given by mutual relations of these building
blocks.

Given some reticular network of fundamental objects we can postulate from purely
logical properties (in the sense of Wheeler's pregeometry) a N-dimensional cubic
lattice. The difference between Hilbert's axiomatic formulation and our approach
is the following: {\sc Hilbert} \cite{13} presuposes the concepts of points,
lines and surfaces, and the axioms are constructed with these objects. In our
approach the points, lines and surfaces are derived from the relational
character of the lattice, but the logical consecuences of the lattice are
equivalent to the set of axioms in Hilbert's formulation.

For our approach we need only one Axiom for each particular lattice.

{\bf Axiom of tesselation:} Given an indefinite number of objects, each of them is
connected with no less and no more than $2N$ different ones, forming a
$N$-dimensional lattice.

This Axiom is called of tesselation (or saturation) because for each dimension
the points are completely connected and don't admit new connexions.

From this Axiom we can derive the definition of straight line, principal
straight lines, orthogonal and paralell straight lines. \cite{14} 

A {\it path} is the connection between two different points, say, A and B, through 
points that are pairwise neighbours.

The {\it length} of a path is the numbers of points contained in the path, including the 
first and the last one.

A {\it minimal path} is a path with minimal length (in the picture the two paths 
between A and B are minimal). Between two point there can be different minimal 
paths.

\bigskip
\begin{center}
\includegraphics{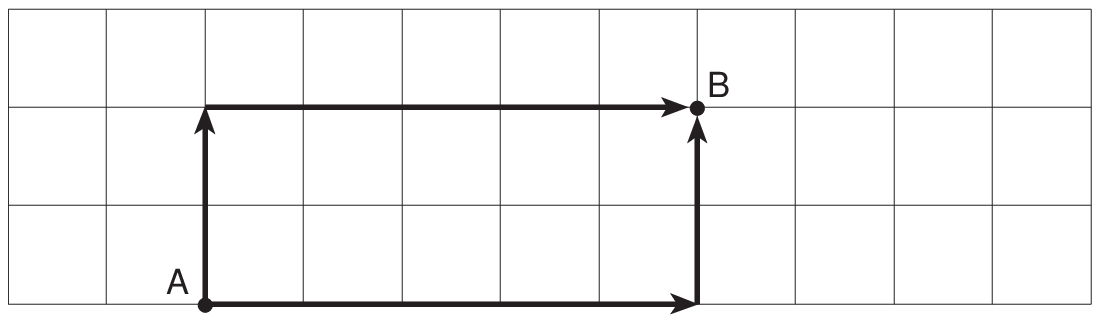}
\end{center}
\medskip

A {\it principal straight line} is a indefinite set of points in the lattice, such that each 
of them is contiguous to other two, and the minimal path between two arbitrary 
points of this line is always unique.

{\noindent {\bf Theorem 1}: through a point of a 2-dimensional square lattice pass only two 
different principal straight lines (they are called {\it orthogonal straight lines}).}

{\noindent {\bf Theorem 2}: two principal straight lines that are not orthogonal have all the
points  either in common or separated (in the last case they are called {\it paralell straight 
lines})}

From these two theorems we can define cartesian (discrete) coordinates and an 
euclidean space. This structure of 2-dimensional space can be easily generalized to
3-dimensional cubic lattice. With the help of cartesian coordinates we can define a (non
principal) straight line passing through two arbitrary points.

It is possible now to deduce from the Axiom of tesselation the properties of the
lattice that are given as Axioms in Hilbert's geometry. For the 2-dimensional
lattice we have:

\begin{enumerate}
\item  There is a unique straight line through any two different points. Any straight line
contains at least two points. There are three non-colinear points.

\item  Given three points A,B,C of a straight line if B is between A and C, A,B,C
are three distinct points, and B is between C and A. Given two points A and C,
there exists one point B over the straight line AC such that C is between A and
B. Given three points on a straight line there is only one between the other two.

\item Given two points A,B of a straight line, and a point $\mbox{A}'$ of
the same or different line there exists a point $\mbox{B}'$ in the same line as $\mbox{A}'$ such
that the segment $\mbox{A}'$$\mbox{B}'$ is congruent with AB. If two segments are
congruent with a third one, they are congruent one to another.

\item  Given a straight line and a point exterior to it, there exists one and only
one straight line that incides in the point and does not cross the first one.

\end{enumerate}

For more complicated lattice we can follow the ``skeleton'' geometry of Wheeler used
to describe the topological properties of curved space-time, that where
developped by Regge and recently by many other authors. \cite{15}

\end{document}